\documentclass[iopjournal,
superscriptaddress,
footinbib,https://www.overleaf.com/project
noeprint,
longbibliography,
prb]{revtex4-2}

\usepackage{amsmath}
\usepackage{listings}
\usepackage{graphicx}
\usepackage{dcolumn}
\usepackage{bm}
\usepackage{graphics}
\usepackage{bbold}
\usepackage{epsfig,color}
\usepackage{dsfont}
\usepackage{mathtools,amssymb}
\usepackage{braket}
\usepackage{stmaryrd}
\usepackage{bbm}

\usepackage[breaklinks,colorlinks,bookmarks=false,citecolor=red,linkcolor=blue,urlcolor=magenta]{hyperref}

\begin{document}
\title{ Machine Learning the  Strong Disorder Renormalization Group  Method for   Disordered Quantum Spin Chains}

\author{ A. Ustyuzhanin}
\email[]{Andrey.Ustyuzhanin@constructor.org}
\affiliation{
Department of Computer Science, Constructor University, Bremen 28759, Germany}

\author{ J. Vahedi}
\email[]{javahedi@gmail.com}
\affiliation{Department of Physics and Earth Sciences and 
Department of Computer Science, Constructor University, Bremen 28759, Germany}

\author{S. Kettemann}
\email[]{skettemann@constructor.university}
\affiliation{Department of Physics and Earth Sciences and 
Department of Computer Science, Constructor University, Bremen 28759, Germany}

\begin{abstract} 

We train machine learning algorithms to infer the entanglement structure of disordered long-range interacting quantum spin chains by learning from the strong disorder renormalization group (SDRG) method. The system consists of $S=1/2$ spins coupled by antiferromagnetic power-law interactions with decay exponent $\alpha$ at random positions on a one-dimensional chain. Using SDRG as a physics-informed teacher, we compare a Random Forest classifier as a classical baseline with a graph neural network (GNN) that operates directly on the interaction graph and learns a bond-ranking rule mirroring the SDRG decimation policy. The GNN achieves a disorder-averaged pairing accuracy of $r_P \simeq 0.94$ and reproduces the entanglement entropy $S(\ell)$ in excellent quantitative agreement with SDRG across all subsystem sizes and interaction exponents. RG-flow heatmaps confirm that the GNN learns the sequential decimation hierarchy rather than merely fitting final-state observables. Finite-temperature entanglement properties are incorporated via the SDRG-X framework through a two-stage strategy—using the zero-temperature GNN to generate the RG flow and sampling thermal occupations from the canonical ensemble—yielding results in agreement with both numerical SDRG-X and analytical predictions without retraining.

\end{abstract}

\maketitle

\section{Introduction}

Magnetic properties of a wide range of materials
are due to randomly placed quantum spins coupled by long-range interactions
\cite{anderson58,loehneysen,Kettemann2023,Kettemann2024,reviewtls,Yu}.
Tunable interactions have been realized in cold atom systems
\cite{grass,Islam2013,Richerme2014,Signoles2021,Brow2020}.
The dynamics of disordered spin ensembles at a diamond surface can be studied
with single nitrogen-vacancy centers \cite{Lukin2022,Davis2022}.
Interacting spin models at finite temperature in a disordered landscape
were simulated in arrays of superconducting qubits, tracking dynamics both in
real space and in Hilbert space \cite{Altman2021,Lunkin2026}.

However, it remains a challenge to model and derive thermodynamic, dynamical,
and entanglement properties of such systems.
The strong disorder renormalization group (SDRG) method was introduced to study
disordered quantum spin chain models
\cite{bhattlee81,bhattlee82,fisher94,fisher95,monthus,Igloi2018}.
Short-range antiferromagnetically coupled disordered spin chains were thereby found
to be governed in the ground state by an infinite-randomness fixed point (IRFP),
whose ground state is a product of randomly positioned singlet pairs
(the random singlet phase).
The SDRG method has been applied to other random quantum spin chains,
such as the transverse-field Ising model \cite{monthus},
and it allows one to derive thermodynamic properties \cite{monthus}
as well as dynamical properties including entanglement dynamics
\cite{Vosk2013,Igloi2012}.

Recently, we applied SDRG to disordered $S=1/2$ spin chains with long-range interactions
\cite{Moure2015,Juhasz2014,Juhasz2016,Moure2018,Mohdeb2020,Kettemann2025},
and compared with numerical exact diagonalization and long-range extensions of DMRG
\cite{Mohdeb2020}.
The SDRG method was extended to study excited states and finite-temperature properties
\cite{Pekker2014,Huang2014,Aram2024} and has recently been applied to study excitations
\cite{Mohdeb2022,Kettemann20252}
and entanglement dynamics after global quantum quenches of long-range coupled spin chains
\cite{Mohdeb2023}.

In parallel, machine learning methods have become increasingly important in quantum many-body physics \cite{Hou23,MehtaReview,RMPReview}.
Artificial neural networks have been used for variational representations of many-body wavefunctions
(neural quantum states, NQS) \cite{Carleo2017,Deng2017,nomura3,choo4,choo11,Rende2025,Viteritti2025,roca1,sharir2},
for a recent review see Ref.~\cite{Lange2024}.
Beyond variational wavefunctions, machine learning has also been used to learn or accelerate
renormalization-group concepts  \cite{Sheshmani23,Mehta,Fan23}  and multiscale transformations, including (i) unsupervised learning of RG
transformations and RG equations \cite{HouYou2023MLRG},
(ii) learning RG-like operator maps targeting a desired property rather than a state ansatz
\cite{Luo2024OLRG},
(iii) learning scale-dependent flow dynamics in functional RG
\cite{DiSante2022DLfRG},
and (iv) learning multiscale, scale-by-scale conditional models with explicit RG structure
\cite{Marchand2022WCRG,ZhaoFoglerYou2025RGFlow}.
Information-theoretic perspectives further support the view of RG as multiscale compression,
including results for disordered systems and inhomogeneous geometries
\cite{Lenggenhager2020OptRG,Gokmen2021RSMI,Gokmen2024Compression,Erdmenger2022RelativeEntropyRG}.

Machine learning has also been applied directly to disordered quantum matter, including
localization physics and disorder-averaged inference
\cite{Ohtsuki2016,Beetar2021MBLGraphs,Chen2024MobilityEdgesCNN}.
Recent work has demonstrated that entanglement-related quantities can be inferred by machine learning
without a full reconstruction of the many-body wavefunction, including GNN-based prediction of
von Neumann entanglement entropy from accessible data \cite{Saleh2025GNNEntropy},
direct machine-learning-based entanglement detection in noise-relevant settings
\cite{Huang2025DirectEntanglement} and quantum information scrambling \cite{Passetti2024DeepLearningScrambling}.
Representation learning and transfer learning can further reduce the cost of learning entanglement
from more accessible observables \cite{Schmidt2024TransferEntropy}.

In this article, we ask whether machine learning models can be trained to infer entanglement properties
of disordered long-range interacting quantum spin systems for a given disorder realization by learning
from SDRG as a physics-informed teacher.
Our aim is not only to reproduce a local decimation criterion, but to approximate the strongly interacting,
nonlocal renormalization flow generated by SDRG in long-range coupled systems, where each decimation step
reshapes effective couplings across many length scales.
Motivated by the success of graph-based policy learning in sequential reduction problems
\cite{Meirom2022RLContraction},
we develop graph-based learning algorithms that take the disorder realization as an irregular weighted
interaction graph and predict entanglement properties without explicit diagonalization.

\begin{it}
Model.
\end{it}—
We consider antiferromagnetic long-range XX-coupled chains of $N$ $S=1/2$ quantum spins,
placed at random positions ${\bf r}_i$ on a line of length $L$, defined by
\begin{equation}\label{H}
H=\sum_{i<j} J_{ij}\left(S_{i}^{x}S_{j}^{x}+S_{i}^{y}S_{j}^{y}\right).
\end{equation}
We take $N$ even with open boundary conditions.
The couplings are antiferromagnetic and decay with a power law in distance
$r_{ij}=|{\bf r}_i-{\bf r}_j|$ with exponent $\alpha$,
\begin{equation}\label{jcutoff}
J_{ij}=J_0\left|({\bf r}_i-{\bf r}_j)/a\right|^{-\alpha},
\end{equation}
with $J_0>0$ and $|{\bf r}_i-{\bf r}_j|>a$ for all $i,j$,
where $a\ll L/N$ is the smallest possible distance.

\section{SDRG-X Method}

We apply  the SDRG-X method,
 as introduced in Ref. \cite{Pekker2014}.
Assuming that many body Eigen states of the Hamiltonian Eq. (\ref{H}) can be written 
 as tensor products of pair states, one  identifies the strongest coupled pair of spins $(i,j)$ for a given
    random positioning of the $N$ spins on the $L$ lattice sites, yielding
    with  the couplings Eq. (\ref{jcutoff}) an 
 initial distribution of couplings $P(J,\Omega_0),$
 where $\Omega_0$ is that largest energy scale. 
We rewrite the Hamiltonian Eq. (\ref{H}) as $\hat{H} = \hat{H}_0 + \hat{V},$
where $ \hat{H}_0 $ is   the Hamiltonian of the most strongly coupled 
 pair of spins  $(i,j),$ given by $\hat{H}_0  =J^x_{ ij}\left(S_{i}^{x}\,S_{j}^{x}+S_{i}^{y}\,S_{j}^{y}\right)$ and  $\hat{V}$
  is the part of the Hamiltonian modeling the  interaction 
 of spins $(i,j)$
 with   all other $N-2$ spins and between them. 
  $\hat{H}_0,$  has   four eigenstates $|s \rangle,$
    $s \in \{0,1,2,3\}$ with eigenenergies
   $E_{s}.$
Projecting 
the  pair $(i,j)$ on one of these pair states $s,$
 one gets  the effective Hamiltionian of the remaining $N-2$ spins 
 $\hat{H}_{\rm eff},$   to 2nd order 
   in  $\hat{V},$ as given in Ref. \cite{Kettemann20252}.
Thereby one obtains  the  renormalized couplings and local fields for all 
 remaining $N-2$ spins. 
 Even  if the form of the Hamiltionian remains 
   unchanged, 
 the  renormalized couplings   differ from the initial ones, and the distribution function of couplings is changed accordingly to 
 $P_E(J,\Omega_0 - d\Omega),$
  with $\Omega_0 - d\Omega$ the 
 largest energy scale in the reduced system of $N-2$
 spins. $E$ is the total energy of the system.
 Repeating this procedure until all $N$
 spins formed pairs, one finds 
  the distribution of  effective couplings $P_E(J,\Omega)$
  in the limit of $\Omega \rightarrow 0$. This allows one 
  to derive thermodynamic and dynamic properties of the spin chains.
  Instead of using a micro canonical ensemble at energy $E,$
  it is often  more convenient to consider a canonical ensemble 
  at bath temperature $T,$
where any of the 
     pair states $s \in \{ 0,...,3\}$  
      is occupied
     at RG scale  $\Omega$     
     with probability
     \begin{equation} \label{occupation}
     p_{s} (E_{s} (\Omega),T) = \exp[- \beta E_{s}(\Omega \sigma)]/Z(\Omega),
     \end{equation}
     with
     $\beta = 1/(k_{\rm B} T).$    
      $Z(\Omega)= \sum_{s} \exp[- \beta E_{s} (\Omega)]$  is the partition sum.

\begin{figure}
    \includegraphics[width=0.3\columnwidth]{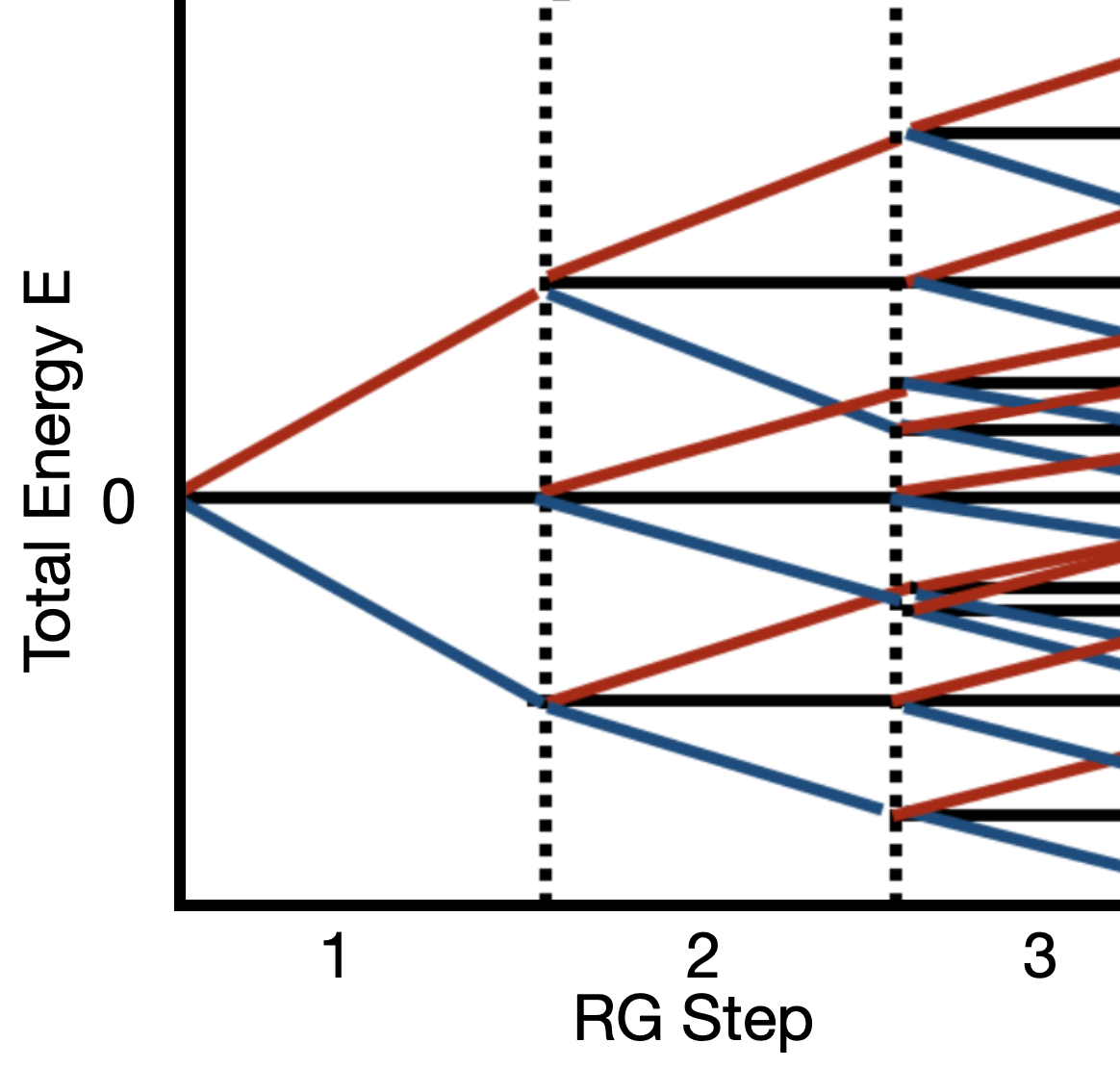} \hspace{1cm}  
    \includegraphics[width=0.5\columnwidth]{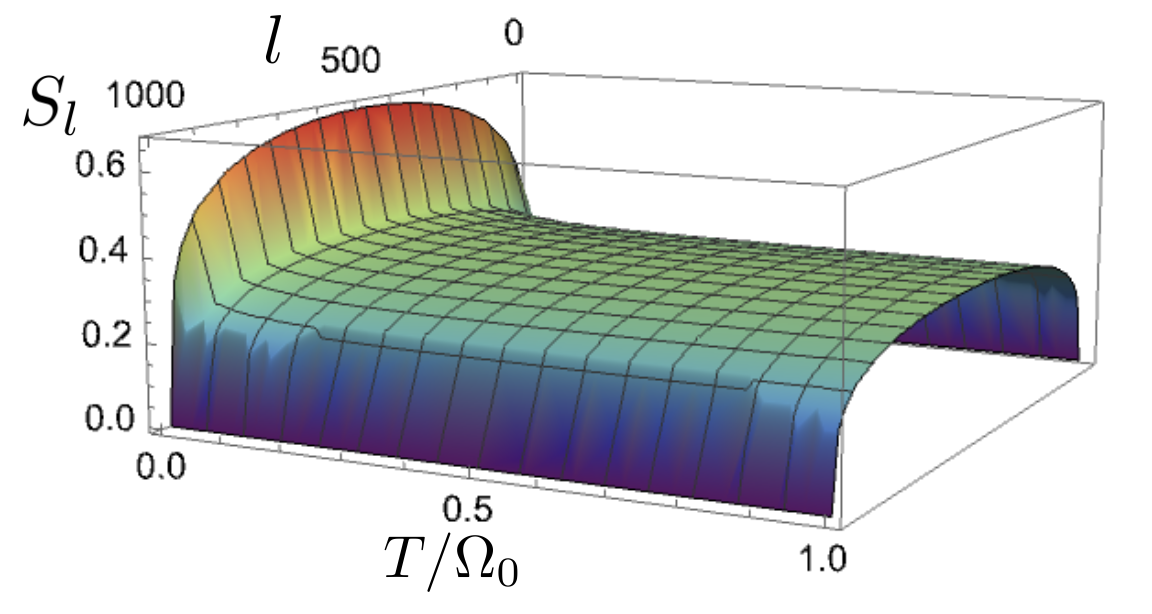}
\vspace*{-.5cm}
\caption{Schematic SDRG-X procedure:
 at each RG step four possible pair states
  are indicated by blue, black and red lines with pair energies $E=-J/2,0,0,+J/2$, respectively. 
  After
  $N/2$ RG steps the  many body eigenstates with total energy E is obtained, following a specific SDRG path. Entanglement entropy as function of physical partition length 
$l$, Eq. (\ref{slL}), valid for $\alpha \gg 1$ for chain length $L=1000$ as function of   temperature $T$ in units of  $\Omega_0$. } 
    \label{RGscheme}
\end{figure}

\section{Entanglement Entropy}

The   distribution function of 
 couplings at finite temperature $T$ has been derived 
  in Ref. \cite{Kettemann20252}, and used to 
 derive thermodynamic and  dynamic properties of bond  disordered spin chains, Eq. (\ref{H}).
  Here, we focus on 
the entanglement entropy of a subsystem 
of physical  length $l$
with the rest of the chain. For a
specific random pair state it is 
given by  $S_l = M_T \ln 2,$ where $M_T$ is the number of singlets ($s=0$) or entangled triplet states ($s=3$)
 at temperature $T$, which are  crossing the partition of the subsystem. 
  The ensemble average of the entanglement entropy is 
 accordingly given by
  $\langle S_l \rangle = 
  \langle M_T \rangle  \ln 2.$
 The average  of $M_T$
 can be derived 
from the distribution of pair  lengths $P_s(l,T).$
For the ground state $T=0,$
 the leading term was found to be  given by\cite{refael},\cite{hoyos},\cite{Mohdeb2020} 
$S_l  \sim \frac{1}{2} \ln 2    \int_{l_0}^n dl' ~ l' P(l').$ 
At finite temperature $T$ we accordingly get 
$S_l  \sim \frac{1}{2} \ln 2    \int_{l_0}^n dl' ~ l' P(l',T) p_T (\Omega_{l'})$,
where $p_T (\Omega_{l'}) = \cosh [\Omega_{l'}/(2T)]/\{1+\cosh [\Omega_{l'}/(2T)]\}$ 
is the probability that the pair with physical  distance $l'$
is in an entangled state ($s=0,3$). Here, 
$\Omega_{l} = \Omega_0 {l}^{-\alpha}.$
 For finite spin chains of length $L,$
  the entanglement entropy at $T=0K$ 
  becomes 
  $S_l (T=0K) \sim \frac{1}{6} \ln 2  \ln [  (L/\pi) \sin ( \pi l/L) ] $ 
  \cite{calabrese}.
At finite temperature $T$ for  $ \alpha  \gg 1$ \cite{Kettemann20252}
and finite length $L$ we find 
  for $\alpha \gg 1$
   \begin{equation} \label{slL}
S_l (T) \sim \frac{1}{6} \ln 2  [ \ln f_L(l) 
-\frac{2}{\alpha} \int_{\Omega_0/(2T f(l)^{\alpha})}^{\Omega_0/(2T)} dz \frac{1}{(1+z)^2\ln z}], 
 \end{equation}
 with $f_L(l) = (L/\pi) \sin ( \pi l/L).$


\section{Teaching Machine Learning Algorithms to Find the Eigenstates of Disordered Quantum Spin Chains}

We investigate how machine learning (ML) algorithms can be trained to reproduce the eigenstate structure of disordered quantum spin chains for a given disorder realization. Our approach is based on learning from the strong-disorder renormalization group (SDRG), which provides an explicit and physically motivated construction of many-body eigenstates in strongly disordered systems. Throughout this section, we consider ensembles of disorder realizations at fixed spin density $N/L=0.1$ and compare ML-assisted results directly with those obtained from  SDRG and its finite-temperature extension, SDRG-X.

To assess the accuracy of the ML-assisted constructions, we employ several complementary measures that probe both microscopic and macroscopic aspects of the eigenstates. At zero temperature, a natural metric would be the wavefunction fidelity
\begin{equation}
F = \bigl|\langle \psi_{\mathrm{ML}} \mid \psi_{\mathrm{SDRG}} \rangle\bigr|^2,
\end{equation}
which directly quantifies the overlap between the ML-predicted state and the reference SDRG eigenstate. However, at finite temperature we work within the canonical ensemble, where each decimated spin pair can occupy any of the four local two-spin eigenstates with finite probability. In this setting, fidelity at the level of the full many-body wavefunction is no longer a meaningful diagnostic.

Instead, we focus on whether the ML algorithm correctly predicts the \emph{pairing structure} generated by the SDRG flow. As a microscopic measure of accuracy, we introduce the pairing ratio
\begin{equation}
\label{rp}
r_P 
= 2\,\frac{M_P}{N},
\end{equation}
where $M_P$ is the number of spin pairs $(i,j)$ that are assigned identically by the ML-assisted procedure and by the reference SDRG construction. This quantity directly probes whether the ML method captures the essential combinatorial structure of the eigenstates, independent of the specific internal quantum state of each pair.

As a macroscopic diagnostic, we compute the entanglement entropy as a function of the subsystem size $\ell$ and compare it with both  SDRG results and analytical predictions,  Eq.~(\ref{slL}). The entanglement entropy provides a stringent test of whether the long-range singlet structure and its finite-temperature modifications are correctly reproduced.

In the following, we apply this evaluation framework to two distinct machine-learning strategies. We first consider a classical approach based on Random Forests, which treats the SDRG decimation step as a supervised classification problem. We then introduce a graph neural network (GNN)–based approach that operates directly on the interaction graph and learns a bond-ranking rule that mirrors the structure of the SDRG flow.

\subsection{Classical baseline: Random Forest}

As a classical machine-learning baseline, we employ a Random Forest (RF) classifier
to learn the SDRG decimation policy from supervised data.
This approach serves as a reference point for the graph-based models introduced
in the following subsection and illustrates the limitations of feature-based
classifiers when applied to strongly nonlocal renormalization-group dynamics.

\paragraph{Data generation and representation.}
Training data are generated by simulating the SDRG-X procedure across multiple
disorder realizations at fixed spin density $N/L=0.1$.
For each realization, $N$ spins are randomly placed on a one-dimensional chain of
length $L \gg N$, and the initial long-range couplings $J_{ij}$ are constructed
according to Eq.~(\ref{jcutoff}) with interaction exponent $\alpha$.
At each RG step, the instantaneous state of the system is represented by the full
effective coupling matrix $J$.
For use with the Random Forest, this matrix is flattened into a feature vector of
dimension $N(N-1)/2$, corresponding to the independent entries in the upper
triangular part of $J$.
To ensure numerical stability, extremely large values are clipped and undefined
entries are set to zero. 

\begin{figure}
    \centering
    \includegraphics[width=0.8\columnwidth]{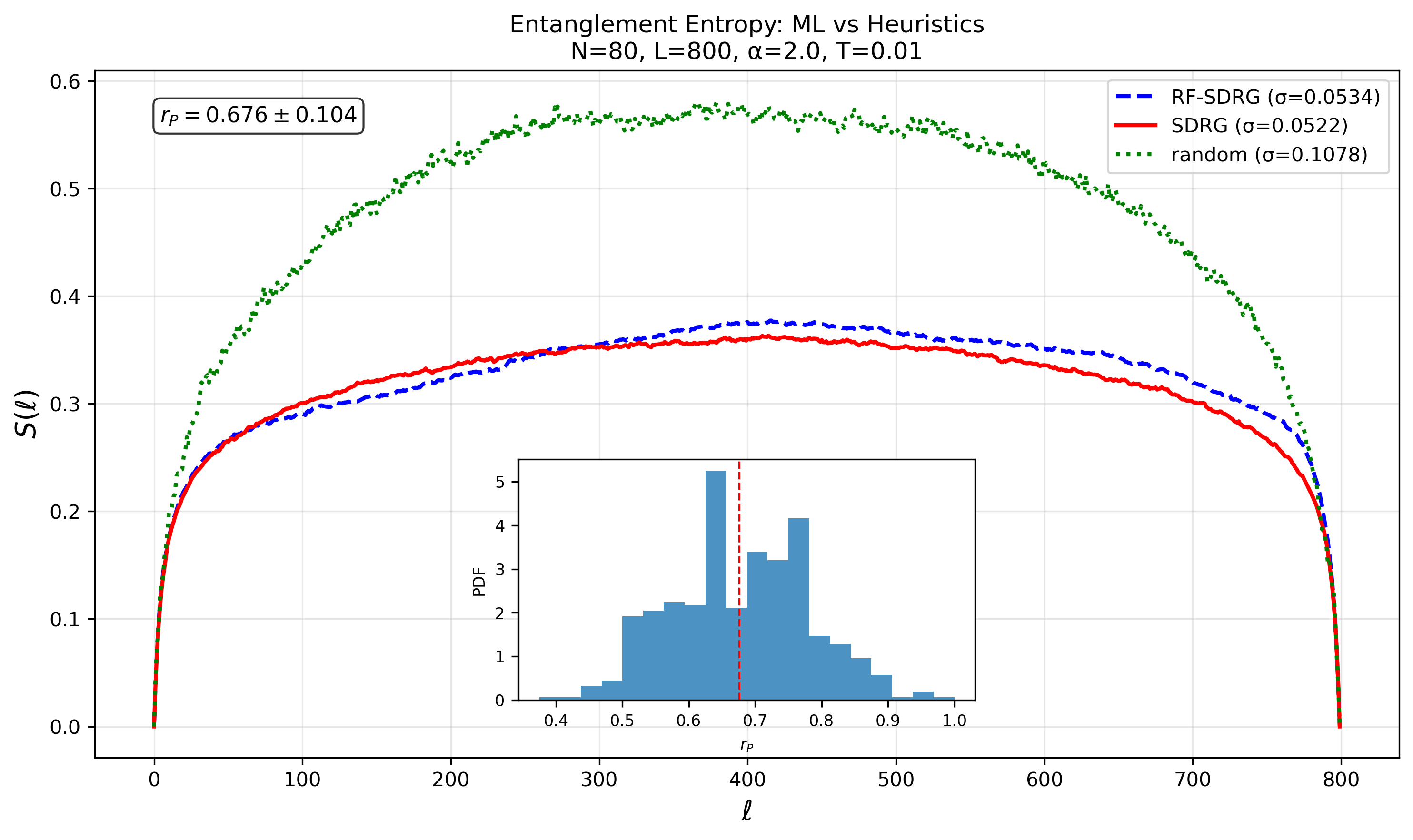}
    \vspace*{-0.5cm}
    \caption{$T=0.01\,\Omega_0$ entanglement entropy $S(\ell)$ as a function of
    subsystem size $\ell$ for a long-range disordered spin chain at fixed density
    $N/L=0.1$, comparing exact SDRG with Random Forest (RF-SDRG) and pure random decimations.
    The top-left annotation reports the mean pairing accuracy
    $r_P = 2M_P/N \pm \sigma_{r_P}$, where $M_P$ is the number of RF-SDRG-predicted
    pairs that match the exact SDRG reference. The bottom inset shows the distribution of $r_P$ across disorder realizations;
    the dashed vertical line marks the mean. All results are averaged over $500$ disorder realizations and $100$ thermal samples per realization.}
\label{MLEE}
\end{figure}

\paragraph{Action space and supervised targets.}
At each RG step, the action consists of selecting a pair of spins $(i,j)$ to be
decimated.
In the flattened representation, this action is encoded as a multi-class label
corresponding to the index of the selected bond in the upper triangular coupling
matrix, resulting in an action space of size $N(N-1)/2$.
Supervised targets are generated using the SDRG strongest-bond criterion, which
selects the pair with the largest effective coupling magnitude.
This choice corresponds to the SDRG decimation rule and provides a well-defined
supervised learning objective.
For benchmarking purposes, alternative baselines such as random bond selection
and coupling-weighted stochastic selection are also considered at evaluation time.

\paragraph{Model architecture and training.}
The Random Forest consists of $n_{\mathrm{trees}}=100$ decision trees, each grown to
full depth using the Gini impurity as the splitting criterion.
At each split, a random subset of features is selected to reduce correlations
between trees and improve generalization.
The model is trained on a dataset comprising approximately
$N_{\mathrm{traj}} \times (N/2)$ samples, where $N_{\mathrm{traj}}$ is the number of
independent disorder realizations and each realization contributes one training
example per RG step.
Training and inference are performed using the \texttt{scikit-learn} implementation.

\paragraph{Inference and ML-assisted SDRG.}
During inference, the trained Random Forest predicts the bond to be decimated at
each RG step.
The predicted class label is mapped back to the corresponding spin pair $(i,j)$,
which is then decimated following the standard SDRG-X rules.
The effective couplings are updated, and the procedure is iterated until all spins
are paired and the RG flow terminates.

\paragraph{Evaluation metrics.}
The Random Forest approach is evaluated using the same microscopic and macroscopic
criteria as the graph neural network models.
At the microscopic level, we assess whether the predicted decimation sequence
reproduces the correct pairing structure obtained from  SDRG, $r_P$.
At the macroscopic level, we compute the disorder-averaged entanglement entropy
\begin{equation}
S(\ell) = \ln 2 \cdot \langle M_T(\ell) \rangle ,
\end{equation}
where $M_T(\ell)$ denotes the number of entangled spin pairs crossing a spatial cut
at position $\ell$.
At finite temperature, entangled pairs include both singlet and entangled triplet
states, sampled according to the canonical ensemble.

\paragraph{Results and limitations.}
Figure~\ref{MLEE} compares the entanglement entropy obtained from SDRG with
that produced by different machine-learning strategies, including the Random Forest
baseline.
While the Random Forest captures qualitative features of the entanglement profile,
substantial quantitative deviations are observed, particularly at larger subsystem
sizes.
These discrepancies highlight intrinsic limitations of the approach.
First, the Random Forest treats each RG step as an independent classification task
and has no explicit notion of the evolving interaction graph or RG history.
Second, the flattened coupling-matrix representation discards permutation
equivariance and spatial structure.
Third, the fixed input dimensionality prevents natural generalization across system
sizes.
Finally, the action space grows quadratically with $N$, making accurate multi-class
classification increasingly difficult for larger systems.
Together, these limitations motivate the use of graph-based models that explicitly
encode interaction structure and multiscale correlations.

\subsection{Graph Neural Network}
To overcome the limitations of classical classifiers such as Random Forests, we introduce a graph neural network (GNN)–based approach that mirrors the structure of the strong-disorder renormalization group (SDRG) itself. Rather than formulating each decimation step as a multi-class classification problem over all possible spin pairs, the GNN learns to assign a relative score to each bond and selects the dominant one via a ranking (pointer) mechanism. This formulation aligns naturally with the SDRG prescription, which is inherently comparative and scale invariant.

\paragraph{Graph representation.}
At each RG step, the instantaneous state of the system is represented as a fully connected graph whose nodes correspond to the currently active spins and whose edges encode the effective couplings between them. Node features are deliberately minimal and consist only of an activity flag, reflecting the fact that SDRG decisions are governed primarily by bond properties rather than by on-site information.

Each edge $(i,j)$ is associated with a feature vector
\begin{equation}
\mathbf{e}_{ij} =
\bigl(
\log |J_{ij}|,
\log |r_i - r_j|,
\log |J_{ij}| - \overline{\log J}_{\mathcal{N}(i,j)}
\bigr),
\end{equation}
where $J_{ij}$ is the effective coupling, $|r_i-r_j|$ is the real-space distance between spins, and the final term measures the bond strength relative to the local neighborhood of spins $i$ and $j$. This local normalization removes explicit dependence on the global energy scale and enforces scale invariance, a defining property of the SDRG flow.

\paragraph{Target and supervision.}
Training data are generated from SDRG trajectories. Each RG step yields a supervised sample consisting of the current interaction graph and a target label identifying the bond $(i,j)$ selected by the SDRG rule, i.e.\ the strongest remaining effective coupling. Unlike the Random Forest approach, the target is not treated as a class label over all possible pairs; instead, it specifies the index of the edge that should receive the highest score.

\paragraph{Dataset generation and preprocessing.}
Supervised training data are obtained by running  SDRG on an ensemble of disorder realizations. For each realization, random spin positions $\{r_i\}_{i=1}^N$ are sampled on a chain of length $L$ with open boundaries, and the initial long-range couplings are constructed as $J_{ij}\propto |r_i-r_j|^{-\alpha}$. The SDRG decimation rule is then iterated until only a single pair remains.

At each RG step, the set of active spins defines a fully connected graph. We store the corresponding node and edge features together with the target edge index identifying the bond selected by SDRG. Each RG step is serialized as an individual sample. Since a realization with $N$ spins produces $N/2-1$ decimation steps, this yields $\mathcal{O}(N\,N_{\mathrm{real}})$ supervised samples. Dataset splits are performed at the level of disorder realizations to avoid leakage of correlated RG steps.

\paragraph{Model architecture and training.}
The GNN consists of an edge-centric message-passing layer followed by a pointer (ranking) head. For each bond $(i,j)$, an embedding is computed by applying a shared multilayer perceptron to the concatenation of node and edge features. A linear readout assigns a scalar score $s_{ij}=f_\theta(\mathbf{e}_{ij})$, and the bond with the largest score is selected for decimation. The model is trained using a cross-entropy loss over edges, optimized with Adam, and evaluated using validation-based checkpointing.

\paragraph{Inference and ML-assisted SDRG.}
During inference, the trained GNN replaces the explicit SDRG decimation rule. At each RG step, edge scores are computed and the highest-scoring bond is decimated. Iterating this procedure yields a full pairing structure analogous to that of  SDRG.

\paragraph{Evaluation metrics.}
Performance is evaluated at both microscopic and macroscopic levels. Microscopically, we measure the accuracy of predicting the correct SDRG decimation at each RG step. To quantify agreement at the level of the final pairing structure, we compute the ratio
$r_P = 2M_P/N,$ and macroscopically, we compute the entanglement entropy $S(\ell) = \ln 2 \cdot \langle M_T(\ell) \rangle,$ where $M_T(\ell)$ is
at $T=0K$
the number of singlet pairs crossing a cut at position $\ell$.

\begin{figure}
    \includegraphics[width=0.8\columnwidth]{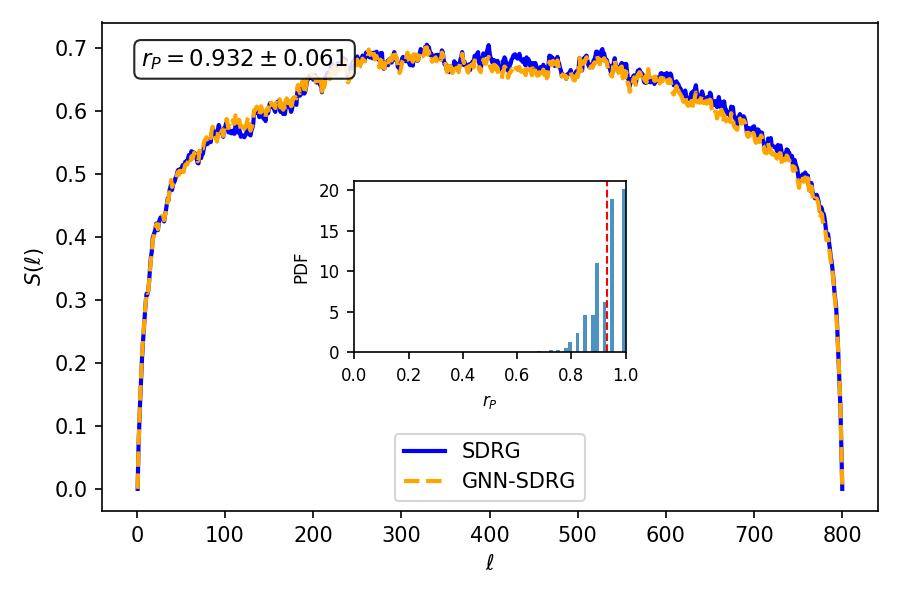}
    \vspace*{-0.5cm}
    \caption{Entanglement entropy $S(\ell)$ at $T=0K$ versus subsystem size $\ell$ for a long-range random spin chain with decay exponent $\alpha = 2.0$ and fixed density $N/L = 0.1$, comparing  SDRG (solid line) and GNN-assisted SDRG (dashed line) results averaged over $1000$ disorder realizations. The inset shows the distribution of the pairing accuracy $r_P$ across disorder realizations; the dashed vertical line indicates the disorder-averaged mean.}
    \label{fig:gnn1_sdrg}
\end{figure}

\begin{figure}
    \includegraphics[width=0.8\columnwidth]{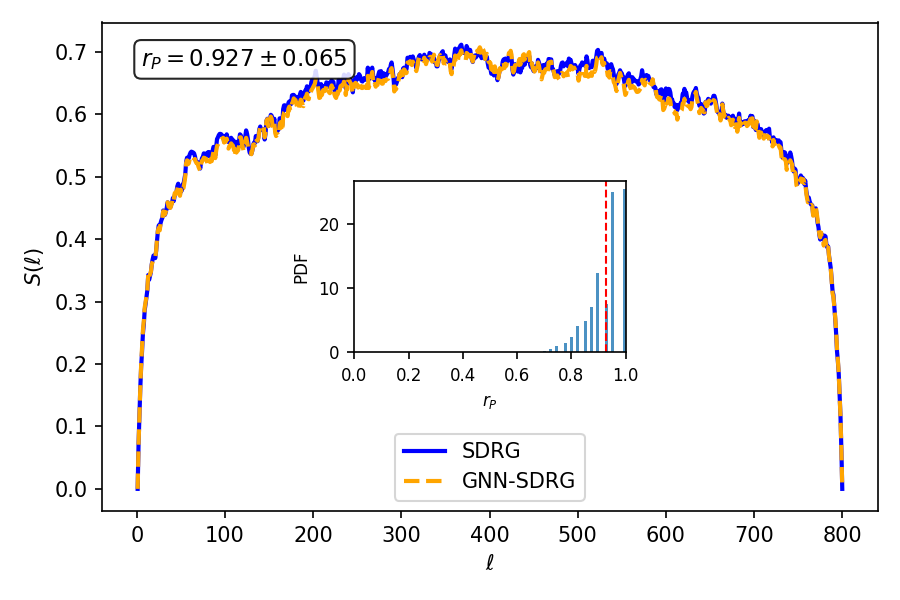}
    \vspace*{-0.5cm}
    \caption{Same as Fig.~\ref{fig:gnn1_sdrg}, but for decay exponent $\alpha = 0.5$ at fixed density $N/L = 0.1$. }
    \label{fig:gnn2_sdrg}
\end{figure}

We begin by assessing the performance of the GNN-assisted SDRG at the level of physically relevant observables.
Figures~\ref{fig:gnn1_sdrg} and \ref{fig:gnn2_sdrg} compare the  $T=0K$ entanglement entropy $S(\ell)$ obtained from  SDRG and from the GNN-assisted SDRG for long-range disordered spin chains at fixed density $N/L=0.1$, for interaction decay exponents $\alpha=2.0$ and $\alpha=0.5$, respectively.

For both interaction ranges, the entanglement entropy profiles produced by the GNN-assisted SDRG are in excellent agreement with those obtained from SDRG across the entire subsystem size $\ell$. The agreement extends from short subsystems to the bulk of the chain and persists up to the boundaries, indicating that the learned decimation policy reproduces the correct accumulation of long-range singlets responsible for entanglement growth.

The insets of Figs.~\ref{fig:gnn1_sdrg} and \ref{fig:gnn2_sdrg} show the distribution of the pairing accuracy $r_P$ across disorder realizations. The distributions are sharply peaked near unity, with a disorder-averaged mean $r_P \simeq 0.94$, demonstrating that the GNN not only reproduces disorder-averaged observables but also recovers the correct pairing structure on a realization-by-realization basis. This high pairing fidelity confirms that the ML-assisted SDRG closely tracks the microscopic pairing decisions of the  SDRG flow.

We note that $r_P$ is a discrete quantity, defined as the fraction of correctly identified singlet pairs out of a finite total number $N/2$. Consequently, the inset histograms exhibit weight only at discrete rational values of $r_P$, resulting in visible gaps between occupied bins. The strong clustering near $r_P\simeq 1$ reflects that the GNN typically makes only a small number of pairing errors, while realizations with multiple correlated deviations from the  SDRG flow are rare. We note tat $r_p$ cannot take the value $1-2/N$ since the state cannot be wrong by a single pair, only.

While agreement of entanglement entropy and pairing structure provides a stringent test of the final SDRG output, it does not by itself guarantee that the GNN captures the renormalization-group dynamics underlying the flow.
To directly probe whether the network has learned the sequential structure of the SDRG, we analyze the distribution of bond lengths decimated at each RG step.

\begin{figure}
    \centering
\includegraphics[width=\columnwidth]{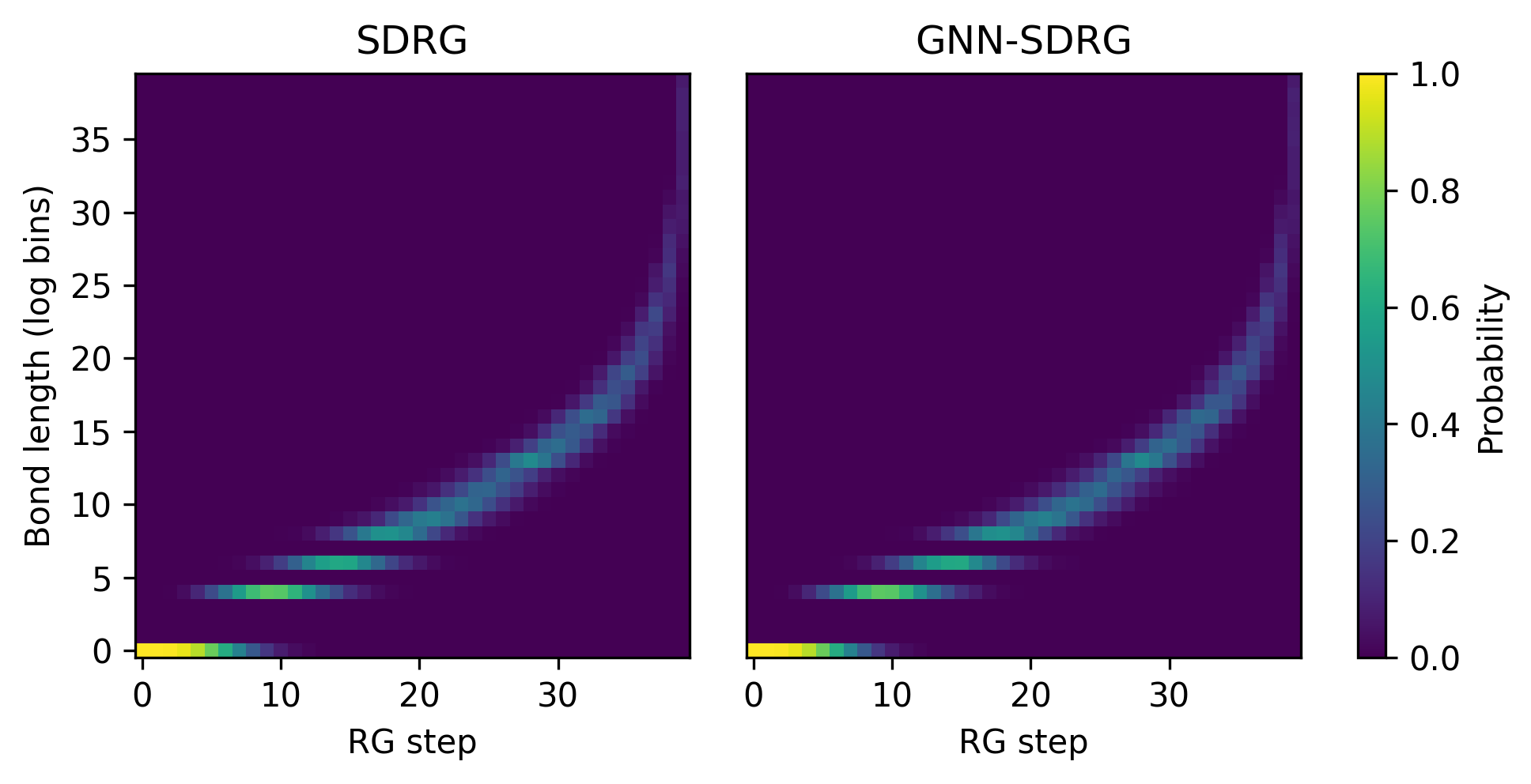}
    \vspace*{-0.4cm}
    \caption{
Renormalization-group flow of bond decimations for SDRG (left) and GNN-assisted SDRG (right) for a long-range disordered spin chain ($N=80$, $\alpha=2.0$).
Shown is a heatmap of the probability of decimating a bond of given length $\ell$ (logarithmic bins) at each RG step, averaged over disorder realizations. }
    \label{fig:rgFlow}
\end{figure}

Figure~\ref{fig:rgFlow} shows the resulting RG-flow heatmaps for SDRG and GNN-assisted SDRG. The vertical axis represents the length of the decimated bond, $\ell = |r_i-r_j|$, binned logarithmically to resolve the wide separation of length scales characteristic of the random-singlet phase. Both methods exhibit the hallmark SDRG hierarchy in which short-range bonds are eliminated at early RG steps, while progressively longer bonds dominate at later stages.

The close agreement between the two heatmaps demonstrates that the GNN reproduces not only final observables such as the entanglement entropy, but also the internal structure of the renormalization-group flow itself. In particular, the network correctly captures the emergence of long-range singlets at late RG times, confirming that it has learned the sequential logic of the SDRG rather than merely fitting its outputs.

\subsection{Finite-Temperature Extension and SDRG-X}

The graph neural network introduced above is trained exclusively on zero-temperature ($T=0$) SDRG trajectories, where the renormalization-group (RG) flow is deterministic and each decimation step is uniquely defined by the strongest remaining bond. This setting yields a well-posed supervised learning problem and allows the GNN to accurately reproduce the $T=0$ random-singlet RG flow. In this regime, the zero-temperature SDRG flow is governed by a universal random-singlet fixed point, for which the entanglement entropy exhibits logarithmic scaling independent of the interaction exponent $\alpha$, up to nonuniversal finite-size corrections.

Finite-temperature physics is incorporated using the SDRG-X framework, which describes thermal ensembles and highly excited eigenstates. A key conceptual feature of SDRG-X is that the \emph{structure of the RG flow remains identical} to the zero-temperature case: at each RG step, the strongest bond is selected based solely on the instantaneous couplings. Temperature enters only at a subsequent stage, through the thermal occupation of the local two-spin eigenstates associated with the decimated bond.

Motivated by this separation, we adopt a two-stage strategy that avoids retraining the GNN at finite temperature. First, the trained $T=0$ GNN is used unchanged to generate the RG flow, i.e.\ the sequence of bond decimations. Second, for each decimated bond with coupling strength $J_{ij}$, the local two-spin eigenstate (one singlet and three triplets) is sampled according to the canonical ensemble at temperature $T$. Finite-temperature observables, such as the entanglement entropy, are obtained by averaging over both disorder realizations and thermal sampling.
 \begin{figure}
    \includegraphics[width=0.8\columnwidth]{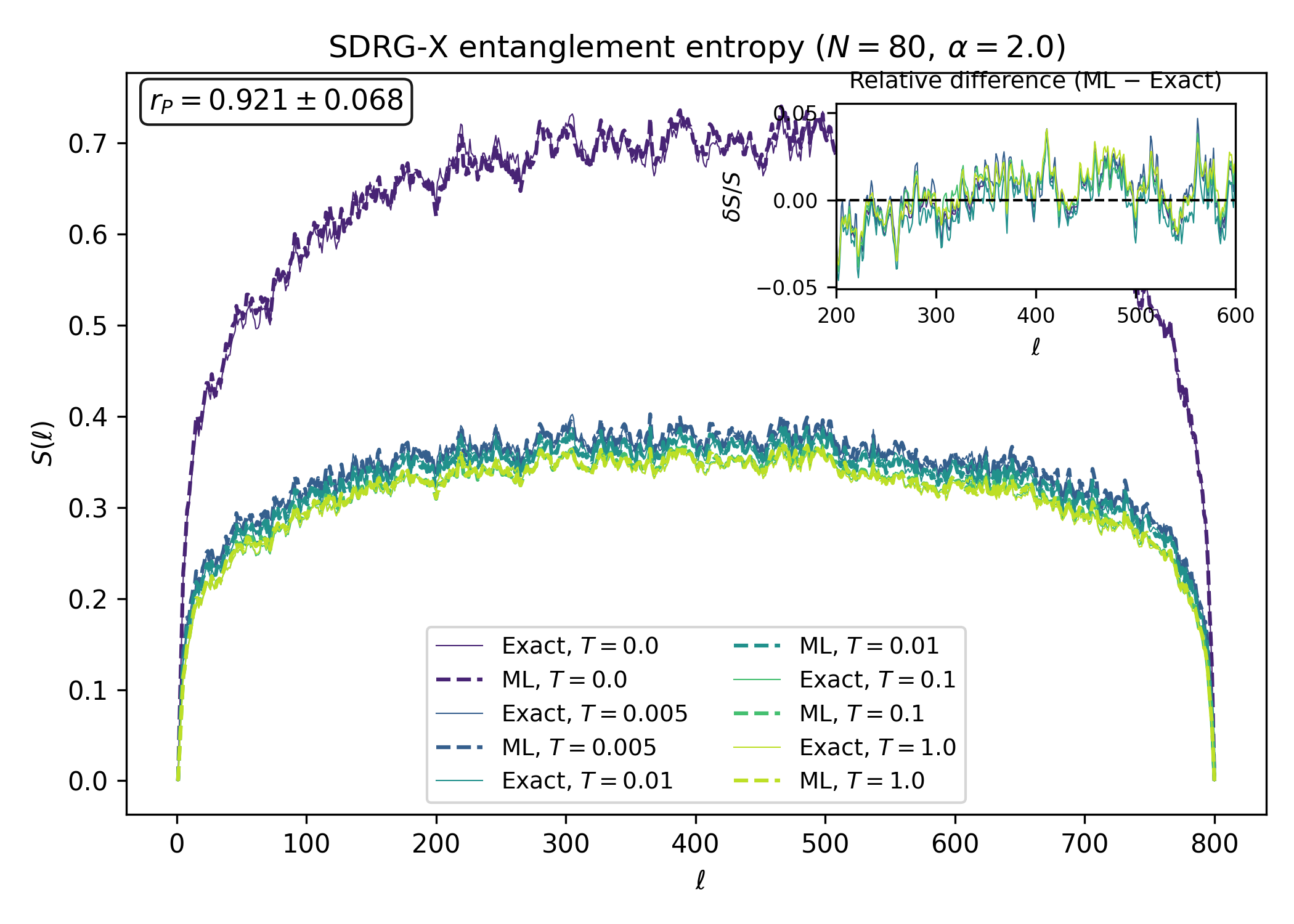}
\caption{
Finite-temperature entanglement entropy obtained from  SDRG-X (solid lines) and ML-assisted SDRG-X (dashed lines) for a long-range disordered spin chain with $N=80$ and interaction exponent $\alpha=2.0$, shown for several temperatures. The GNN is trained exclusively at zero temperature and is used unchanged to generate the renormalization-group flow, while finite-temperature effects are incorporated via SDRG-X thermal sampling.
The main panel shows the entanglement entropy $S(\ell)$ as a function of the cut position $\ell$, demonstrating excellent agreement between SDRG and ML-assisted results across all temperatures. The inset displays the relative difference $(S_{\mathrm{ML}}-S_{\mathrm{SDRG}})/S_{\mathrm{SDRG}}$ in the bulk of the system, which fluctuates around zero and remains within statistical uncertainty.
All results are averaged over $500$ disorder realizations and $100$ thermal samples per realization.
}\label{fig:ml_sdrgx_2.0}
\end{figure}

Alternative machine-learning strategies for finite-temperature SDRG-X can be envisaged. One possibility would be to train a GNN directly on finite-temperature SDRG-X trajectories, with stochastic labels reflecting the thermal occupation of local eigenstates.
While such an approach could in principle learn both the RG flow and thermal probabilities simultaneously, it would introduce intrinsic label noise and render supervised training ill posed.
Another option would be to extend the network output to predict both the decimated bond and the corresponding eigenstate occupation.
We do not pursue these directions here, as the present two-stage strategy leverages the deterministic structure of the RG flow and incorporates thermal effects in a controlled and physically transparent manner.

Figure~\ref{fig:ml_sdrgx_2.0} shows the resulting finite-temperature entanglement entropy $S(\ell)$ for several temperatures, comparing SDRG-X with ML-assisted SDRG-X based on the GNN-generated RG flow. For all temperatures considered, the two results are visually indistinguishable on the scale of the main plot, indicating that the learned decimation policy faithfully captures the RG structure governing long-range entanglement. The inset displays the relative difference $(S_{\mathrm{ML}}-S_{\mathrm{SDRG}})/S_{\mathrm{SDRG}}$ in the bulk of the chain, which fluctuates around zero and remains small compared to statistical uncertainties from disorder and thermal sampling.

While finite-temperature entanglement observables acquire an explicit $\alpha$ dependence through the thermal cutoff of the RG flow, the underlying bond-selection structure learned by the GNN remains governed by the same zero-temperature universality. This approach cleanly separates the learning of the temperature-independent RG structure from the incorporation of thermal physics via known statistical mechanics. It avoids stochastic supervision during training, allows temperature to be varied \emph{a posteriori}, and yields finite-temperature observables that are directly comparable to analytical and numerical SDRG-X results.

 \begin{figure}
    \includegraphics[width=0.8\columnwidth]{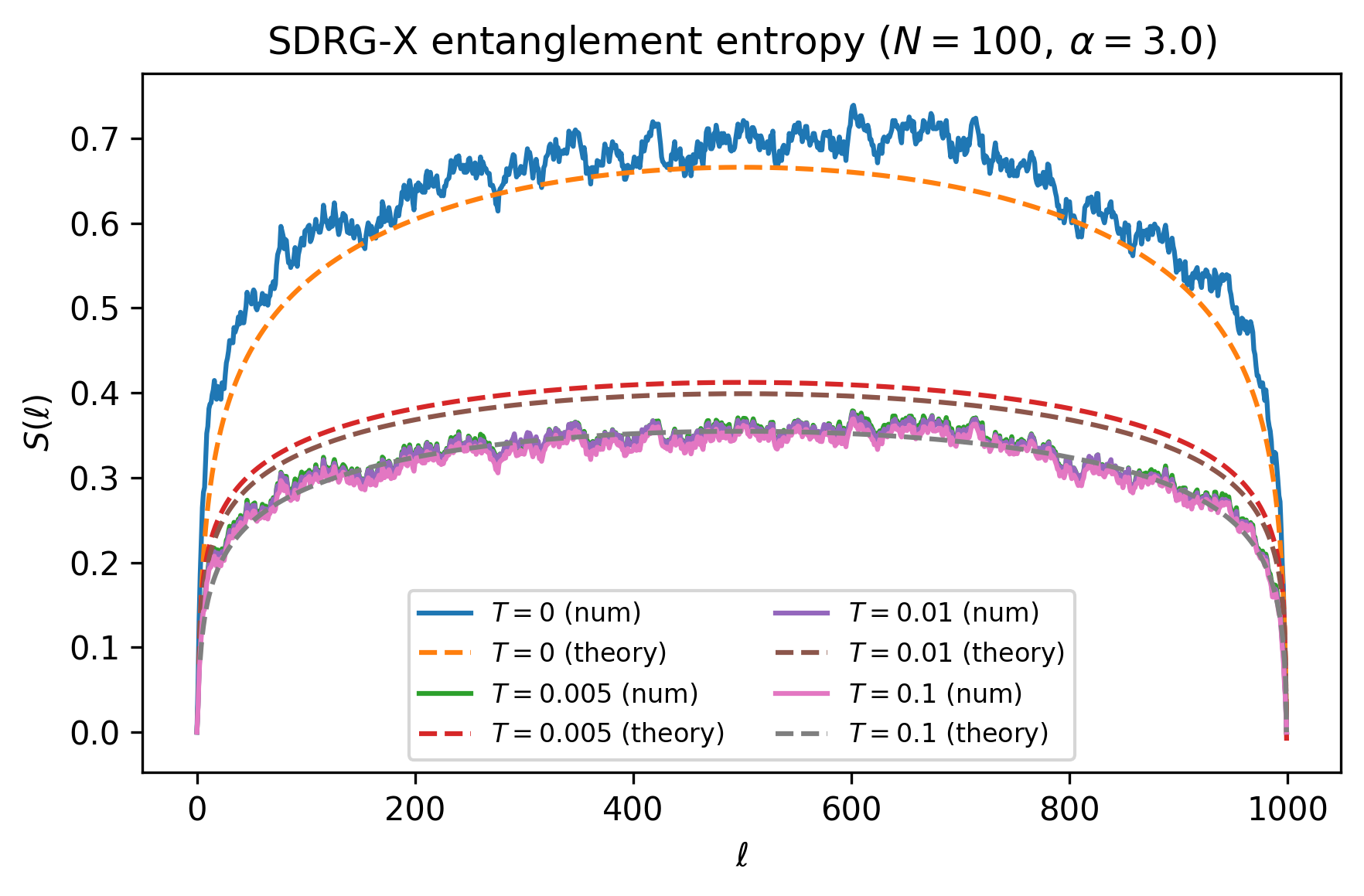}
\vspace*{-.5cm}\caption{
Entanglement entropy $S(\ell)$ obtained from SDRG-X for a long-range interacting spin chain with $N=100$ and $\alpha=3.0$.
Solid lines show numerical results averaged over $500$ disorder realizations and $200$ thermal samples, while dashed lines represent the analytical prediction of Eq.~(\ref{slL}) for finite temperature and finite system size.
}
\label{sdrgx_2.0}
\end{figure} 

To further validate the finite-temperature behavior captured by the SDRG-X framework, we compare the numerical results directly with analytical predictions.
Figure~\ref{sdrgx_2.0} shows the entanglement entropy $S(\ell)$ obtained from  SDRG-X for a long-range interacting spin chain with $N=100$ and $\alpha=3.0$, together with the analytical finite-temperature expression of Eq.~(\ref{slL}) evaluated at finite system size.
At zero temperature, the numerical data follow the random-singlet finite-size form, while at finite temperature the entropy is lower since there is a finite probability that spin  pairs are in one of the two unentangled triplet states. 
Overall good agreement is observed at large $\ell$, confirming that SDRG-X correctly captures the  finite-temperature scaling of the entanglement entropy.

\subsection{Conclusions}

We  teach machine learning algorithms to find 
the entanglement properties of disordered quantum spin systems for a given disorder realization 
  by learning 
from the strong disorder renormalization group method, as applied to a training set of  disorder realizations. 


The SDRG flow generates a hierarchical pairing structure in which short-range bonds are eliminated first and long-range singlets emerge only after successive coarse-graining steps. Capturing this sequential and scale-invariant structure is essential for reproducing the correct entanglement scaling. Classical machine-learning approaches that ignore the RG flow, such as random or weighted pairing rules, fail to incorporate these constraints.

 The graph neural network performs therefore better than rain forest,  because it learns the renormalization-group structure underlying the pairing process, rather than merely fitting final-state observables. This suggests that ML models trained on RG data may provide a powerful and transferable framework for studying more complex disordered quantum systems, including higher-dimensional lattices and models with more intricate interaction graphs, where explicit RG constructions become increasingly challenging
\cite{Hikihara2020TensorSDRG2D,Seki2021EntanglementSDRG,Hikihara2023AutoStructTTN,Ozawa2022RGNNTopological}.
 
\section{Acknowledgments}

    {Acknowledgments.- J.V. acknowledges the Ataro Group for generously providing office space and computational resources that facilitated the completion of this work.

\end{document}